# Towards Autonomic Service Provisioning Systems

Michele Mazzucco
*Department of Computer Science*
*University of Cyprus*

*Abstract*—This paper discusses our experience in building SPIRE, an autonomic system for service provision. The architecture consists of a set of hosted Web Services subject to QoS constraints, and a certain number of servers used to run session-based traffic. Customers pay for having their jobs run, but require in turn certain quality guarantees: there are different SLAs specifying charges for running jobs and penalties for failing to meet promised performance metrics. The system is driven by an utility function, aiming at optimizing the average earned revenue per unit time. Demand and performance statistics are collected, while traffic parameters are estimated in order to make dynamic decisions concerning server allocation and admission control. Different utility functions are introduced and a number of experiments aiming at testing their performance are discussed. Results show that revenues can be dramatically improved by imposing suitable conditions for accepting incoming traffic; the proposed system performs well under different traffic settings, and it successfully adapts to changes in the operating environment.

## I. Introduction

In the last few years one of the main challenges for information technology has been the integration of applications within and across organizational boundaries. The use of Web Services eases the interoperability between different systems because it allows programs to interact with each other over the Internet via open protocols and standards like SOAP and HTTP. One of the main concerns, however, is that users and customers may face major problems and eventually incur major costs if computing systems do not meet the expected performance requirements: customers expect reliability and performance guarantees, while under-performing systems loose revenues. For example, Google found that an extra 0.5 seconds in search page generation would kill user satisfaction, with a consequent 20% traffic drop [1], while trimming the page size of Google Maps by 30% resulted in a traffic increase of 30% [2]. Also, it has been reported that Amazon tried delaying the page generation by 100 ms and found out that even very small delays would result in substantial and costly drops in revenue (1% sales drop for 100 ms delay) [3]. Thus, as Web Services proliferate more and more widely, whether offered within an organization or as part of a paid service across organization boundaries, the issues related to service quality become very relevant and will eventually be a significant factor in distinguishing the success or the failure of service providers (75% of shoppers who have a poor experience on a Web site will not shop on that site again [4]). Over-provisioning is an expensive solution, and it is not guaranteed to work under extreme conditions such as flash-crowds. On the other hand, even with the adoption of the data center as the hub of IT organizations and provider of business efficiencies the problems are not over, as it is extremely difficult for service providers to meet the promised performance guarantees in spite of unpredictable traffic. One possible approach is the adoption of Service Level Agreements (SLAs), contracts that specify a level of performance that must be met and compensations in case of failure. Quality of Service (QoS) issues can be addressed from several points of view, such as the engineering point of view (*i.e.*, how to provide a service subject to performance constraints), or the semantic one (*i.e.*, how to dynamically discover or select services with tight performance requirementsIn this paper we focus on the first one. More in detail, we investigate a very important but often neglected aspect of web-based systems, the cost of service provision. From the provider's perspective, the problem can be defined as *'How to maximize the earned revenue?'*, that is, how to minimize the probability of failing to honour the commitments for agreed service quality. In order to do that, we use a combination of admission control algorithms, service differentiation, resource allocation techniques and economic parameters to make the service provisioning system as profitable as possible. As far as we are aware, there is no existing commercial system that corresponds to the model described here, while previous studies focus only on some of those techniques (see Section VI).

### A. Contributions and Paper Organization

The main contributions of this paper are:
1) The formalization of infrastructure technologies to determine costs and penalties for users and providers;
2) The design and implementation of a middleware solution that allows a service provider to control and structure a commercial data center running session-based traffic subject to SLAs;
3) The implementation of autonomic algorithms aiming at improving the efficiency of service provisioning systems by allowing them to adapt to changing demand conditions.

The rest of this paper is organized as follows. Section II presents the background. Section III introduces the system model, discusses the properties autonomic provisioning systems should have, and presents the SLAs and the utility functions controlling the system. Section IV describes the system we have built, while a number of experiments are reported in Section V. Section VI discusses relevant related work, while Section VII concludes the paper.

## II. BACKGROUND

This section introduces the core idea of SPIRE (Service Provisioning Infrastructure for Revenue Enhancement), a management system for enterprise data centers designed with a utility computing paradigm in mind.

Currently, most of the data centers are operating under stringent performance requirements. These requirements can be either dictated by users, if the data center is hosting some Web applications, or can be stated in Service Level Agreements (SLAs), if the server farm is offering a service to paying customers. When designing an enterprise hosting platform different architectures can be employed, the most widely ones being *shared* and *dedicated*. As the name implies, the shared architecture does not allocate entire servers. Instead, it runs multiple applications on each server and multiplexes the server resources among these applications. On the other hand, the dedicated model is a hosting platform where servers are not shared: each application runs on a subset of the available servers, while each machine is allocated to at most one application. If the contract that regulates the service provision allows the host to use the same server to run different applications, then a dynamic allocation scheme can be used. This means that, no matter what definition of load is used, the service provider can periodically estimate the load for each application and change the amount of machines running the hosted services accordingly. During the design phase, we decided to follow the dedicated pattern as it eases the management of the hosting infrastructure. In particular, in SPIRE changing the number of resources running a certain application is simply a matter of routing requests: if the system is carefully designed (see Section IV), this task is extremely flexible and can be performed very quickly.

## III. SYSTEM MODEL

A central challenge in the management of commercial data centers is the necessity to keep them continuously optimized. Utility functions provide a simple way for achieving self-optimization, as systems driven by utility functions use appropriate optimization techniques to determine the most valuable feasible state and the way to reach that stage.

### A. Requirements

In order to to be successful, we require from our system:
1) To be capable to simultaneously manage different services with different service quality requirements;
2) To be able to effectively react to changes in user demand;
3) To be light-weight, *i.e.*, to be able to adjust its behavior using traffic estimates only;
4) To provide protection against overload conditions, as without proper protection throughput drops under heavy load, while waiting and response times grow to unacceptable levels;
5) To be able to cope with session-based traffic, *i.e.*, group of interactions – see the formal definition below – as it is a very important class of traffic (*i.e.*, eBay, Amazon), but requires special solutions.

In order to achieve these goals, SPIRE uses a combination of admission control algorithms, service differentiation, resource allocation techniques and economic parameters to make the service provisioning system as profitable as possible. Dealing with session-based traffic requires ad-hoc algorithms, as job-based admission control policies drop requests at random [5]. Therefore all clients connecting to the system would be likely to experience connection failures or broken sessions under heavy load, even though there might be capacity on the system to serve all requests properly for a subset of clients. Moreover, since active sessions can be aborted at any time, there could be an inefficient use of resources because aborted sessions do not perform any useful work, but they 'waste' system resources.

*Definition 1 (Session):* A session of type $i$ is a collection of $k_i$ jobs, submitted at a specified rate of $\gamma_i$ jobs per second.

Our model requires session integrity (*i.e.*, if a session is accepted, all jobs in it will be executed) as it is a critical metric for commercial Web Services. From a business perspective, the higher the number of completed sessions, the higher the revenue is likely to be, while the same does not apply to single jobs. Apart from the penalties resulting from the failure to meet the promised QoS standards, broken sessions or delays at some critical stages, such as checkout, could mean loss of revenue for the service owners. From a customer's point of view, instead, breaking session integrity would generate a lot of frustration because the provided service would appear as not reliable.

### B. QoS, SLA and Utility Functions

In order to guarantee an adequate level of performance, the service provider and the clients are bound by an SLA specifying that the QoS experienced by an accepted session $i$ is measured by the observed average waiting time, $W_i$. Such a value should not be greater than a specified threshold, otherwise the provider is liable to pay a penalty back to the customer. $W_i$ is computed as the arithmetic mean of the waiting times of the jobs belonging to that session (*i.e*, the interval between a job's arrival and the start of its service). One could also decide to measure the QoS by the observed average response time, taking also the job lengths into account. The question of whether to use response time or waiting time as the QoS measure is largely one of marketing. From the point of view of the provider, waiting time is better because there is less uncertainty associated with it (lower variance), and the admission policy is simpler. On the other hand, users might prefer an SLA based on response times.

Each SLA agreement includes, among the others, three clauses specifying the charge the customer must pay for having his/her jobs run, the obligations of the service provider, and the penalty that the provider has to pay if he/she fails to meet the promised service quality. In this paper we implement and experiment with three different reward functions.

*1) Flat penalties:* The first utility function uses a 'flat penalty' factor for the penalties [6], *i.e.*, the provider pays the penalty specified into the SLA, no matter what the amount of

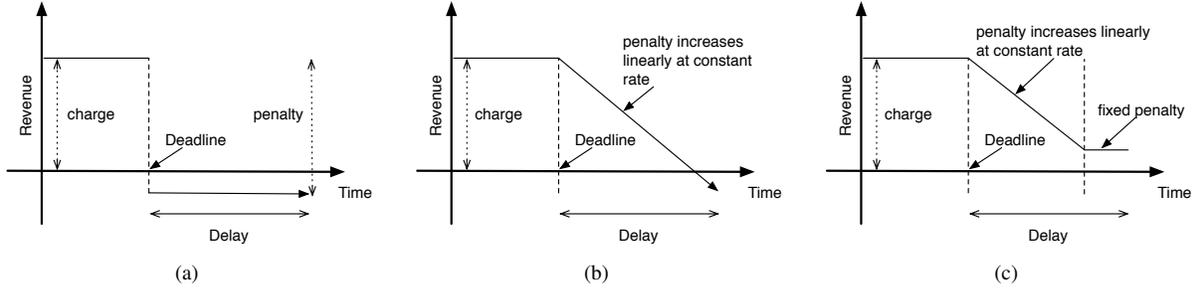

Fig. 1. Utility function with (a) flat penalties, (b) penalties proportional to the delay, and (c) penalties proportional to the delay with bound penalty.

the delay is (see Figure 1(a)). The SLA for this model defines charges, obligations and penalties in the following terms:

*Definition 2 (Flat penalty):* For each accepted session of type $i$ whose average waiting time exceeds the obligation (*i.e.*, $W_i > q_i$), the provider is liable to pay to the user a penalty of $r_i$.

*Definition 3 (Charge):* For each accepted session of type $i$, a user shall pay a charge of $c_i$.

How to determine the amount of the charge is outside the scope of this paper. However, intuitively this could depend on the number of jobs in the session and their submission rate, or on the obligation (*i.e.*, higher charges for stricter obligations).

*Definition 4 (Obligation):* The observed average waiting time, $W_i$, of an accepted session of type $i$ shall not exceed $q_i$.

While the performance of computing systems can be measured using different metrics, in this paper we are interested in the average revenue earned by the service provider per unit time, $R$, as it is more meaningful from a business perspective than values such as the throughput (*i.e.*, people running business activities are usually interested in profit, not in throughput or response time values). Having defined charges, obligations and penalties, $R$ can be computed using the following expression:

$$R = \sum_{i=1}^{m} a_i[c_i - r_i P(W_i > q_i)], \qquad (1)$$

where $a_i$ is the average number of type $i$ sessions accepted into the system per unit time.

*2) Penalties proportional to the delay:* The realism of the previous model can be disputed, as a delay exceeding the obligation of 0.1 second would generate the same amount of penalty as a delay 5 or 50 times greater than the promised performance quality. In order to overcome this limitation, we introduce a new reward function, where the amount of penalty the service provider is liable to pay to the user is proportional to the amount of delay experienced by the accepted sessions, see Figure 1(b). In order to do so, the penalty is re-defined as follows:

*Definition 5 (Proportional penalty):* For each accepted session of type $i$ whose average waiting time exceeds the obligation (*i.e.*, $W_i > q_i$), the provider is liable to pay to the user a penalty of $r_i \times (W_i - q_i)$.

The corresponding utility function is the following:

$$R = \sum_{i=1}^{m} a_i[c_i - r_i \times max(0, W_i - q_i)], \qquad (2)$$

*3) Proportional penalties with upper bound:* The second model can be very punitive for the service provider, thus we propose an extension that limits the maximum amount of penalty that the provider is liable to pay:

*Definition 6 (Proportional penalty with upper bound):* For each accepted session of type $i$ whose average waiting time exceeds the obligation (*i.e.*, $W_i > q_i$), the provider is liable to pay to the user a penalty of $r'_i \times (W_i - q_i)$ if $q_i < W_i \leq t_i$, and a fixed penalty of $r''_i$ for higher delays.

The new reward function is:

$$R = \begin{cases} \sum_{i=1}^{m} a_i[c_i - r'_i \times max(0, W_i - q_i)] & \text{if } W_i \leq t_i \\ \sum_{i=1}^{m} a_i[c_i - r''_i] & \text{if } W_i > t_i \end{cases}$$
(3)

This model differs from the previous because the delay is now proportional to $W_i$ only for delays in the interval $q_i + \epsilon, \ldots, t_i$, as in Equation (2), and fixed to $r''_i$ for $W_i > t_i$ (as in Equation (1), see Figure 1(c)).

*Remarks:* Please note that the formulae used to compute the utility functions (2) and (3) differ from the one used by Equation (1), as the system needs to estimate the expected average waiting time for a given set of traffic parameters, and not the probability of exceeding the promised threshold. Also, if this methodology is to be applied in a real hosting environment, it may be necessary to carry out some market research in order to find out which SLA the users would be more willing to subscribe, what kind of waiting time obligations they might ask for, and how much they would be willing to pay for them.

Possible extensions to this model include:
- The cost function may include the cost of switching servers from one job type to another;
- Instead of allocating whole servers to job types, one could share servers between different job types, but control the fraction of service capacity each type uses (*i.e.*, via block of threads). If that is the case, those fractions would play the role of servers, eventually with different state-dependent service times.

## C. Self-Managing Policies

The random nature of user demand and changes in demand pattern over time make capacity planning very difficult in the short time period and almost impossible in the long time period. It is clear that in such situations it could be advantageous to reallocate resources from one type to another, even at the cost of switching overheads. The question that arises in that context is how to decide whether, and if so when, to perform such system reconfigurations. Posed in its full generality, this is a complex problem which not always yields an exact and explicit solution. For this reason, SPIRE embeds some heuristic policies which, even though not optimal, perform reasonably well and are easily implementable.

During the intervals between consecutive policy invocations, the number of active sessions (or the maximum number of active sessions, if the 'Threshold' policy is used) remains constant. Such intervals, or *observation windows*, are used by the controlling software to monitor the traffic and collect statistics in order to estimate values such as arrival rates ($\lambda_i$), service times ($b_i$), and associated variability ($ca_i^2$ and $cs_i^2$). Such estimates are then used to perform the queueing analysis at each configuration epoch in order to make decisions about server allocations and sessions admission during the next window. It is worth noting that all of the above parameters are time varying and stochastic in nature, and thus their values should be estimated at each configuration interval. However, if the estimates are accurate enough, the arrival rates and service times can be approximate as independent and identically distributed (i.i.d.) random variables inside each window, thus allowing for online optimizations.

The idea of using windows for self-optimizing computing systems is not new, but while SPIRE uses event-based windows, other scientists propose time-based windows [7] or complex algorithms to detect traffic surges [8]. The use of events, instead, provides a simpler way to implement *adaptive windows*: under heavy traffic conditions the allocation and admission algorithms are executed more often than when the load is light. The core idea behind adaptive windows is that there is no need to be very clever when there is no traffic, as all requests will be accepted, while the system should perform close to the optimum under heavy load.

In this paper, we implement and experiment with various heuristic policies. As allocation algorithm we propose the 'Offered Loads' heuristic (see Fig. 2), a simple adaptive policy that, using the traffic estimates collected during the previous observation window, allocates the servers roughly in proportion to the offered loads, $\rho_i = \lambda_i b_i$, and to a set of coefficients, $\alpha_i$, reflecting the economic importance of the different jobs types (for service differentiation purposes).

For admission purposes, SPIRE embeds two heuristics, 'Current State' and 'Threshold'. These algorithms are formally described in [6], and thus we only summarize them here. The 'Current State' policy estimates, at every arrival epoch, the changes in expected revenue, and accepts the incoming session (possibly in conjunction with a reallocation of servers from other queues) only if the change in expected revenue is positive. In order to compute that value, it uses the state of each queue, which is specified by the number of currently active sessions, the number of completed jobs and average waiting time achieved so far (for each session).

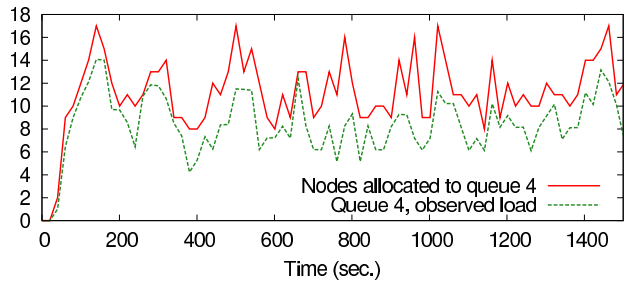

Fig. 2. Dynamic resource allocation. Resources are allocated in proportion to the measured load.

The 'Threshold' heuristic uses a threshold, $M_i$, for each job type, and an incoming session is accepted into the system only if less than $M_i$ sessions are active. Each threshold $M_i$ is chosen so as to maximize the revenue of queue $i$, $R_i$ (for a given allocation, the different services can be decoupled and considered in isolation of each other). We have carried out some numerical experiments, and found that $R_i$ is a unimodal function of $M_i$. That is, it has a single maximum, which may be at $M_i = \infty$ for lightly loaded systems. That observation implies that one can search for the optimal admission threshold by evaluating $R_i$ for consecutive values of $M_i$, stopping either when $R_i$ starts decreasing or, if that does not happen, when the increase becomes smaller than some $\epsilon$. Such searches are typically very fast.

## IV. DESIGN CHALLENGES

The overall increase in traffic on the Internet causes a disproportional increase in user demand to popular Web sites, especially in conjunction with special events. System administrators continually face the need to increase the server capacity. An easy approach would be to mirror the information across the available servers. Unfortunately such model is not transparent to the users because they should manually choose a URL. Besides, it does not provide any load balancing mechanism. A better solution would consist in a distributed architecture capable of routing the jobs among the available servers in a flexible and transparent manner. This model can be implemented in different ways, for example ($i$) via DNS, ($ii$) through a dispatcher or ($iii$) via a two-level dispatching mechanism involving the DNS as well as the servers in the cluster. While all of these alternatives have pros and cons [9], SPIRE uses the dispatcher pattern as it is the most flexible approach: as shown in Figure 3, the dispatcher hides the IT infrastructure from the clients and creates an illusion of a single system by using a Layer-7 two-way architecture [10]. The load balancer ($i$) forwards packets in both directions, client-to-server and server-to-client (packet double-rewriting), and ($ii$) takes routing decisions using only the information available at the application layer of the OSI stack, such as target URL or cookie.

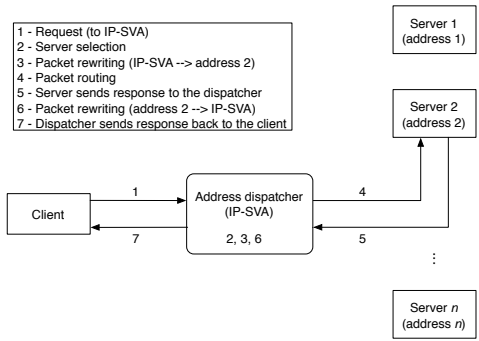

Fig. 3. Packet double-rewriting model: the dispatcher has a single, virtual IP address (IP-SVA) and rewrites packets in both directions (steps 3 and 6).

Using this mechanism it becomes straightforward to add or remove servers, even at runtime, because clients do not know where their requests will be executed. The main advantage of Layer-7 over Layer-4 load balancers is that Layer-7 load balancers offer good policy management functionality, *i.e.*, the ability to define, integrate with existing policies and enforce, during the runtime cycle, policies such as access control and SLA compliance. Layer-4 load balancers, instead, perform essentially a content-blind dispatching, which is faster and easy to implement, but less efficient because the employed routing algorithms are essentially stateless.

*A. Architecture Overview*

Today's service provisioning systems are usually designed according to a three-tier software architecture. The first one translates end-user markup languages such as HTML or XML into and out of business data structures, the second tier (*i.e.*, the business logic tier) performs computation on business data structures while the third level provides storage functionality. Requests traverse tiers via synchronous communication over local area networks and a single request may revisit tiers more than once. Business-logic computation are often the bottleneck for Internet services, and thus we focus on this tier. However, user-perceived performance depends also on disk and network workloads at other tiers. Front-end servers are not typically subject to a very high workload, and thus over-provision is usually the cheapest solution to meet service quality requirements. Moreover, different solutions exist to address some of the issues occurring at both the presentation and database tiers (*e.g.*, [11] and [12]), while [13] has shown that smart scheduling can improve the performance of the database tier.

A high level view of the resulting architecture, which has successfully been deployed at British Telecom's R&D Laboratories where it is used for experimental purposes, is depicted in Figure 4. SPIRE dynamically groups the available machines into virtual pools, where each pool deals with demands for a particular service (*i.e.*, it uses a dedicated hosting model, see discussion in Section II). All client requests are sent to the Controller (arrow 1), which performs all resource allocation, admission control (notifying users of rejections, arrow 7) and monitoring functions. For each type of service there is a corresponding Service Handler, which implements the scheduling policy (arrow 2), the collection of statistics through a profiler, and the management of the currently allocated pool of servers. If the admission policy does not require global state information (*e.g.*, threshold-based policies), then it too may be delegated to the Service Handlers. Each Service Handler also keeps track of the active sessions, which are stored into a hash-table, and updates session-related statistics as arrival and completion events occur. If the same service is offered at different QoS levels (*e.g.*, gold, silver and bronze) and a threshold-based admission policy is employed [6], the Service Handler will be instantiated at differentiated service levels. Each differentiated level will have its own SLA management function instantiated that strives to meet that level of service specified by the differentiation. If the load is too high for any of the differentiated services, then the admission policy will start rejecting incoming traffic in order to maintain an adequate level of performance. For state-based policies (*i.e.*, policies that take into account the state of all queues at every decision epoch), instead, there is no need to use different Service Handlers to deal with different QoS levels, as sessions can specify their own QoS requirements. The advantage of this approach is that it reduces the fragmentation during server allocations. Finally, the results of completed jobs are returned to the Controller, where statistics are collected, and then to the relevant user (arrows 5 and 6).

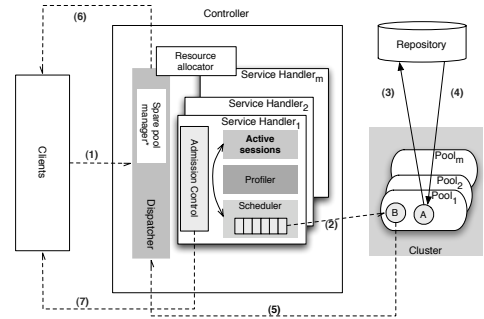

Fig. 4. Architecture overview. Dotted lines indicate asynchronous messages.

If a deployment is needed, the service is fetched from a remote repository (arrows 3 and 4). This incurs a migration time of a few seconds, while a server is switched between pools. SPIRE tries to avoid unnecessary deployments, while still allowing new services to be added at runtime. In such cases, the appropriate Service Handlers are automatically created in the Controller. If there is sufficient space on a server, deployed services could be left in place even when not currently offered on that server. One would eventually reach a situation where all services are deployed on all servers. Then, allocating a server to a particular service pool does not involve a new deployment; it just means that only jobs of that type would be routed to it. In those circumstances, switching a server from one pool to another does not incur any overhead.

It is worth emphasizing that no assumption is made about the number of machines running the Controller. The application can be scaled-out using common techniques (*i.e.*, group communication, session-replication, etc.) in order to prevent the Controller from becoming a bottleneck or a single point of failure. Also, techniques like the one proposed in [14] can be employed in order to operate a cluster on a single IP address while avoiding network congestion on the Controller.

*B. Implementation of the Mediation Service*

This section discusses the implementation details of the controller. The prototype has been implemented in Java and relies on the Apache Axis2 framework[1] to handle SOAP messages. Even though SPIRE is transport agnostic, it is assumed that the parties communicate using the HTTP protocol, as this is by far the most widely used protocol to exchange SOAP messages over the Internet. The message forwarding algorithm uses the WS-Addressing[2] information available in the SOAP header of incoming messages to take routing decisions and to dispatch them: SPIRE uses a custom Axis2 handler placed on the controller's chain in order to change the addressing information of the incoming messages and redirect them to the mediation service. The concept of message interceptor (or handler) is a widely used concept in messaging systems: its task is to intercept the messaging flow and do whatever task it is assigned to it, such as message validation or content enrichment.

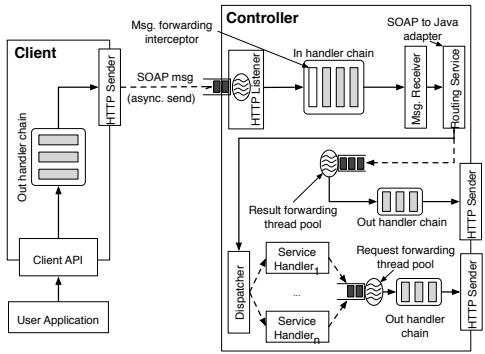

Fig. 5. Axis2-based implementation of the Layer-7 load balancer.

Figure 5 illustrates the internals of the load balancer. The component handing incoming messages (jobs and results as well as internal messages and new sessions) is built on top of an event-driven layer, where the messaging processing pipeline is divided into multiple stages separated by queues, as it happens in the Stage Event Driven Architecture (SEDA) [15]. The HTTP listener initializes the execution of incoming messages. Before reaching the target service every message passes through the input sequence of interceptors. The first handler of the input chain is the interceptor performing the message forwarding: the handler gets the SOAP header and retrieves the message type. Only messages coming from other SPIRE components (servers and repository) have a header portion specifying the type of message the system is trying to handle. In other words, if there is no message type, the received message is a client request.

After passing through the interceptors' pipe, requests are handled by the mediation service, called `RoutingService`, which is implemented as a stateful Web Service. It is essentially an intermediate layer used to validate the XML information contained into the SOAP header of the received messages, to translate them into Java objects, and to forward the results back to the client. It provides operations to add new machines and services at runtime, and to handle requests and responses. All the operations exposed by the `RoutingService` are defined as one-way operations, even though the client expects a response: the message rewriting algorithm will take care of delivering the result of the computation to the client as soon as it becomes available. The mediation service works at the XML level, *i.e.*, it does not need any WSDL nor intermediate layers to marshal and unmarshal the message content, and it uses exclusively the information contained into the message header to take routing and admission control decisions. This implies, for example, that SPIRE can handle encrypted messages without any problem.

## V. EXPERIMENTS

*A. Testbed Settings*

Several experiments were carried out in order to evaluate the robustness of SPIRE. As mentioned in Section III, the metric of interest is the average earned revenue per unit time. CPU-bound jobs (we want to stress the business logic tier, see Section IV-A) whose lengths and arrival instants were randomly generated, queued and executed. We use synthetic load as it allows us to easily vary the distribution of service and interarrival times. Moreover, it let us abstract from the hardware details such as number of cores, clock rate or amount of memory; this way a job takes the same time everywhere, no matter on which hardware it is executed. Apart from the random network delays, messages are subject to random processing overhead, which cannot be controlled. The server capacity is guaranteed by a cluster of 20 (identical) servers running Linux with kernel 2.6.14, Sun JDK 1.5.0_04, Apache Axis2 1.3 (to handle SOAP messages) and Tomcat 5.5 (to handle HTTP requests). The connection between the load generator and the controller is provided by a 100 Mb/sec Ethernet network, while the servers of the cluster are connected to the controller via a 1 GBit/sec Ethernet network. The average round trip time (RTT) between nodes and the controller is 0.258 ms, while the one between the client and the controller is 0.558 ms. Nevertheless, since both the servers and the network are shared, unpredictable delays due to other users are possible. Random inter-arrival intervals were generated by client processes while service times were randomly generated at the server nodes. Each server can execute only one job at any time, *i.e.*, the system does not allow processor sharing (in

---

[1] http://ws.apache.org/axis2/.

[2] WS-Addressing provides transport-neutral mechanisms to address and route messages, see http://www.w3.org/Submission/ws-addressing/.

Section VII we suggest the possibility to extend the current system by running multiple jobs concurrently in a controlled way in order to maintain the same QoS guarantees). The scheduling policy is FIFO, with no preemption, while servers allocated to job type $i$ cannot be idle if there are jobs of type $i$ waiting. Jobs of type $i$ arrive at rate $\lambda_i$, while required service times have mean $b_i$. The following parameters are kept fixed:

- The QoS metric is the average waiting time, $W$;
- The number of jobs in each session, $k$, is 50;
- The average service time, $b$, is 1 second;
- The SLA states that the maximum average waiting time of jobs will be less than or equal to their average service time, *i.e.*, $q_i = b_i$.

Four services whose settings are summarized in Table I were deployed on SPIRE. The total offered load, $\rho$, is increased by varying the submission rate for sessions of type 4, $\delta_4 \in (0.02, \ldots, 0.2)$. At the lower end this represents a 60% loaded system, whereas at the higher end, if all sessions were accepted, the system would be over-saturated, as the total load would be 105%.

| Index | $b_i$ | $\gamma_i$ | $\delta_i$ | $\lambda_i = k_i \delta_i$ | $\rho_i = \lambda_i b_i$ |
|---|---|---|---|---|---|
| 1 | 1.0 | 2.0 | 0.10 | 5.0 | 5.0 |
| 2 | 1.0 | 2.0 | 0.04 | 2.0 | 2.0 |
| 3 | 1.0 | 2.0 | 0.08 | 4.0 | 4.0 |
| 4 | 1.0 | 1.0 | 0.02 … 0.2 | 1.0 … 10 | 1.0 … 10 |

TABLE I
EXPERIMENT SETTINGS. $m = 4, N = 20$.

### B. Performance Evaluation

*1) Flat Penalties:* The first experiment, shown in Figure 6(a), measures the average revenues obtained by the heuristic policies when all charges and penalties are the same, *i.e.*, $c_i = r_i = 10, \forall\ i$: if the average waiting time exceeds the obligation, users get their money back. For comparison, a policy that accepts all incoming traffic – 'Admit all' – is also displayed. Each point in the figure represents a SPIRE run lasting about 2 hours. In that time, between 1,400 (low load) and 1,700 (high load) sessions of all types are accepted, which means that about 70,000 – 85,000 jobs go through the system. Samples of achieved revenues are collected every 10 minutes and are used at the end of the run to compute the corresponding 95% confidence interval (the Student's $t$-distribution was used). The most notable feature of this graph is that while the performance of the 'Admit all' policy becomes steadily worse as soon as the load increases and drops to 0 when it approaches the saturation point (see Figure 6(b)), the heuristic algorithms produce revenues that grow with the offered load. According to the information we have logged during the experiments, they achieve that growth not only by accepting more sessions, but also by rejecting more sessions at higher loads.

The second result concerns a similar experiment, except that now charges and related penalties differ between each job type: $c_1 = 10$, $c_2 = 20$, $c_3 = 30$ and $c_4 = 40$. Moreover, if the SLA is not met, users get back twice what they paid, *i.e.*, $r_i = 2c_i$. Intuitively, when the penalties are larger than the charges, making the right admission and allocation decisions assumes a greater importance. The figures to come make this clear, by illustrating the increasing difference between using a good policy and not using one. As shown in Figure 6(c), now the revenues obtained by the 'Admit all' policy become negative as soon as the load starts increasing because penalties are very punitive. The 'Threshold' policy performs very well if compared to the other heuristics, as it is a very conservative policy (*i.e.*, it is much more likely that it rejects a session instead of violating an SLA).

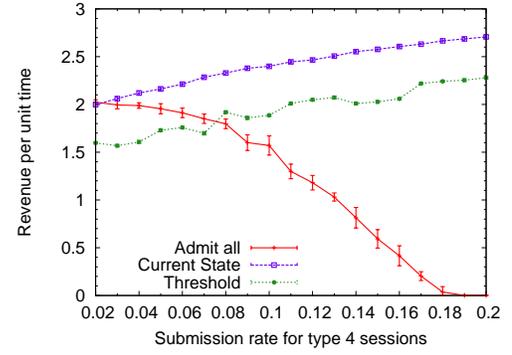

(a) $c_i = r_i = 10, \forall\ i$.

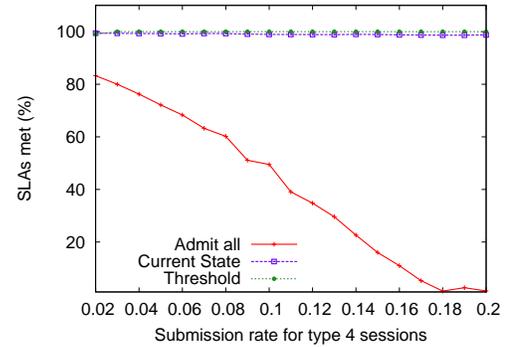

(b) SLAs met for different policies.

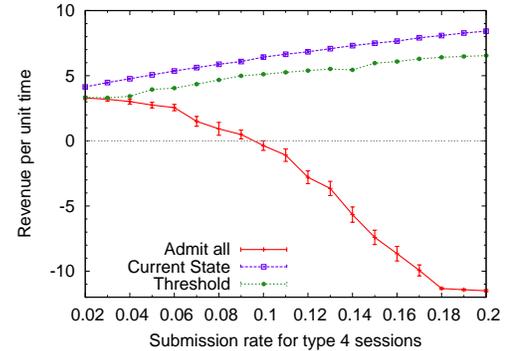

(c) $r_i = 2c_i$.

Fig. 6. Observed revenues for different policies: Markovian scenario, flat penalties.

Next, the Poisson job arrival processes are replaced with bursty arrivals. More precisely, if the overall arrival rate for jobs of a given type is $\gamma$, then 80% of the inter-arrival intervals are on the average $1/(5\gamma)$, and 20% are $4.2\gamma$. This increases the squared coefficient of variation of inter-arrival times, $ca^2$, to 6.12. The aim of increasing variability is to make the system less predictable and decision making more difficult. It is worth stressing that it is not the session submission rate, $\delta_i$, the one subject to bursts, but the arrival rate, $\gamma_i$, at which jobs belonging to a session arrive. When the penalties are higher than the related charges, like in Figure 7(a), the increased unpredictability caused by the bursty arrivals have some effect on the obtained revenues compared to the Markovian case (see Figure 6(c)), with the 'Threshold' heuristic performing almost as well as the 'Current State' policy. The shape of the revenues is similar to the previous try, however the revenues obtained by the 'Current State' heuristic are now about 20% smaller than in the Markovian case.

The previous experiments tested the behavior of the admission control policies under different conditions. However, the loading conditions did not change over the time, while usually the volume of demand in production application environments fluctuates on several time scales (*i.e.*, daily and monthly cycles). Thus, the robustness of SPIRE is assessed under non-stationary traffic conditions. The total load is the same as before, *i.e.*, $\rho$ ranges between 60% and 105% by varying the rate at which type 4 sessions arrive to the system, however, every $x$ seconds, $\delta_1$ and $\delta_2$ are swapped. In other words, during period 1, $\delta_1 = 0.1$ and $\delta_2 = 0.04$, during period 2 $\delta_1 = 0.04$ and $\delta_2 = 0.1$, and so on. As consequence, the loads for types 1 and 2 fluctuate between 2 and 5. The other parameters are the same as in Figure 6(a): $c_i = r_i = 10$, $\forall i$, while the service time and inter-arrival interval distributions are exponentially distributed. For comparison reasons, the figures include an 'Oracle Threshold' policy, that knows exactly when the load variations occur and recomputes the allocation and admission threshold vectors accordingly. In Figure 7(b), the session submission rates change every 300 seconds.

Because the allocation and admission decisions are taken every time a session arrives or completes, the 'Current State' algorithm is not affected by the changes in $\delta_1$ and $\delta_2$, and so it is capable to perform as well as when the load is stationary (see Figure 6(a)), while the revenues obtained by the 'Threshold' policy are between 72% and 83% of the ones achieved under constant load. The main reason is that the 'Threshold' policy uses traffic estimates to compute the vector of admission thresholds: the 'Oracle' does it by using the $\delta_i$ values, while the ones that periodically recompute the two vectors estimate the traffic parameters as well as their variability. As discussed earlier, this policy is very conservative, and its behavior is further emphasized by the fact that the steady state is never reached. From a practical point of view, the 'Threshold' algorithm does not seem very sensible with respect to the employed window size (we also experimented with $x = 60$ seconds and $x = 600$ seconds and obtained similar results): using a bigger window size allows to system to achieve a slightly higher revenue (close to the 'Oracle'), but at the expense of a bigger confidence interval.

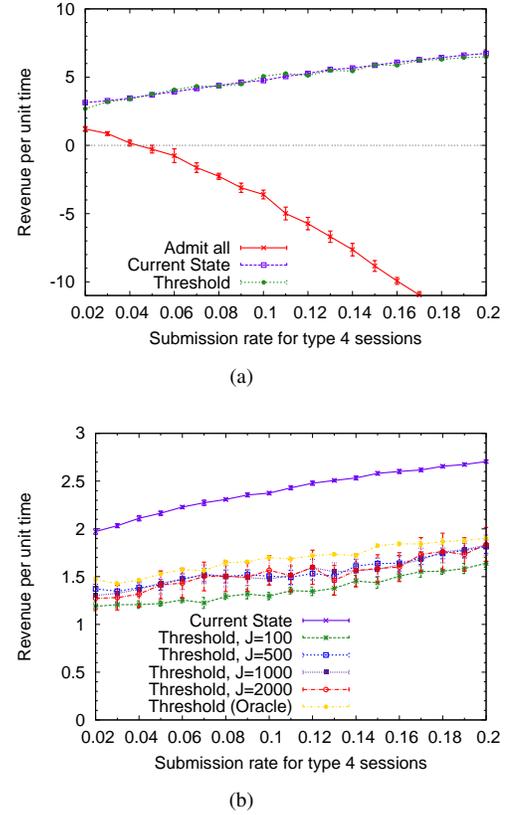

Fig. 7. (a) Bursty scenario, $ca^2 = 6.12$, $r_i = 2c_i$, and (b) Variable load ($\rho_1$ and $\rho_2$ swapped every 300 seconds), $c_i = r_i = 10$, $\forall i$.

*2) Penalties proportional to the delay:* Next, we validate the model using penalties proportional to the experienced delay. Figure 8(a) measures the average revenues obtained by the two heuristics when $c_1 = 10$, $c_2 = 20$, $c_3 = 30$ and $c_4 = 40$, while the base penalties (*i.e.*, the penalty that the provider has to pay when $W_i = q_i + \epsilon$) are $r_i = c_i/2$. Like in Figure 6 jobs enter the system according to independent Poisson processes, while service times are exponentially distributed with means $b_i$.

As before, while the performance of the 'Admit all' policy becomes worse and worse as the load increases, the heuristic policies produce revenues that grow with the offered load. Moreover, since the system is unstable for $\delta_4 = 0.2$, under 'Admit All' the number of queued jobs (and thus the waiting/response times) grows unbound, as shown in Figure 8(b). Since the penalties are proportional to the delay, they keep growing too: for longer lasting experiments the revenue produced by the 'Current State' policy would be the same, while the loss produced by the 'Admit All' policy would be bigger! Finally, the simple 'Threshold' heuristic does not perform as well as the 'Current State' because it behaves in a very conservative way, thus rejecting much more sessions and missing potential revenues: the former rejects 55% of the incoming sessions, while the latter only 35%.

*3) Penalties proportional to the delay with upper bound:* Finally, we test the reward function that limits the maximum

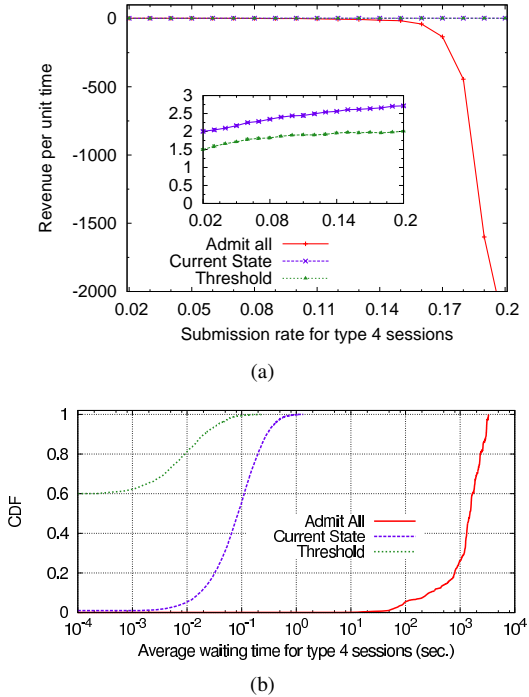

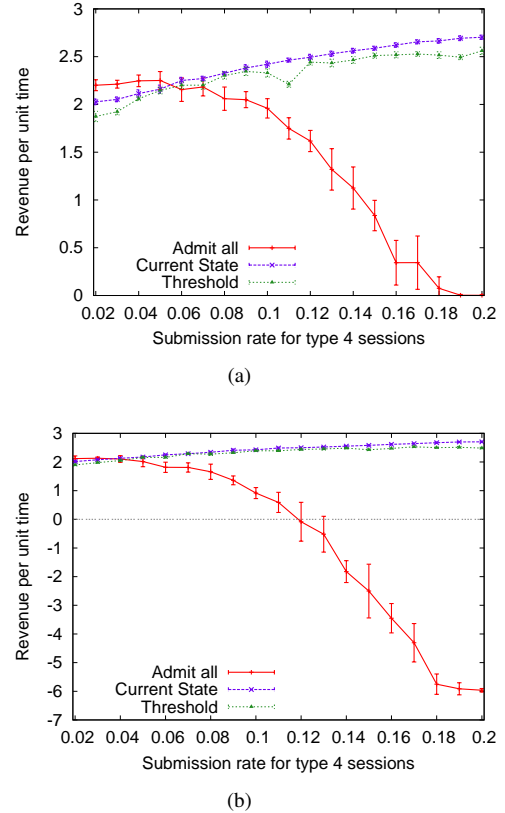

Fig. 8. Penalties proportional to the experienced delay: (a) Observed revenues for different policies, and (b) Cumulative Distribution Function (CDF) of the delays experienced by type 4 sessions for $\delta_4 = 0.2$.

Fig. 9. Observed revenues for different policies, penalties proportional to the experienced delay, with upper bound set to (a) $t_i = 2q_i, r_i'' = 2r_i$, and (b) $t_i = 5q_i, r_i'' = 5r_i$, and base penalty $r_i' = r_i/2$.

amount of penalty that the provider is liable to pay in case of poor service quality. The traffic is still Markovian, while we vary the value of the parameters $t_i$ and $r_i''$ in Equation (3). In Figure 9(a) they are both set to 2, while in Figure 9(b) they are set to 5. The most notable result is that while in the first experiment the 'Admit All' policy achieves no losses under heavy load, as the charges and the penalties are the same (the base penalties, $r_i$, are set to $c_i/2$), in the second scenario the revenues obtained by the policy that does not reject any incoming session become negative, even though they are bound. On the other hand, the performance of the two heuristics are similar, and they do not seem too much affected by the change in the parameter values.

## C. Summary

The experiments we have carried out show that, in order to make the system as profitable as possible, a smart allocation policy is not enough. Instead, it should be combined with an admission policy in order to meet the promised performance quality under heavy load. Moreover, the system should be able to automatically adapt to changes in the operating environment. Various throttling algorithms have been tested and shown to perform well under different traffic conditions. According to the experiments, it seems that the 'Threshold' heuristic would be a good candidate for practical implementation – the 'Current State' heuristic performs slightly better, but is more demanding in terms of computational overhead [6].

## VI. RELATED WORK

Our approach uses a combination of admission control algorithms, service differentiation, resource allocation techniques and economic parameters to make the service provisioning system as profitable as possible. As far as we are aware, there is no existing commercial system that corresponds to the model described here. There is an extensive literature on adaptive resource management techniques for commercial data centers (*e.g.*, [16], [17]). However, previous work does not take into account the economic issues related to SLAs. As a consequence the service providers would still need to over-provision their data centers in order to address highly variable traffic conditions. Moreover, existing studies do not consider admission policies as a mechanism to protect the data center against overload conditions, while we have shown that admission control algorithms have a significant effect on revenues. Several studies (see, for example, [18] and [19]) use workload profiling to estimate the resource savings of multiplexing workloads in a shared utility. Such studies focus on the probability of exceeding the agreed performance requirements for various degrees of CPU overbooking. Others vary the degree of overbooking to adapt to load changes, but they do so by considering only average-case QoS within each interval [12]. [16] considers a resource allocation model for QoS management, where application needs may include timeliness, reliability, security and other non functional requirements.

The model is described in terms of a utility function to be maximized and is further extended in [17]. However, although those schemes allow for variation of job computation time and frequency of application requests, once again congestion and response or waiting time constraints are not considered.

The problem of autonomously configuring a computing cluster to satisfy SLA requirements is addressed in several papers. Some of them consider the economic issues occurring when services are offered as part of a contract, however they do not address the problems affecting overloaded server systems (*e.g.*, [20], [21]), while others include simple admission control schemes without taking any economic parameter into account when the system configuration needs to be updated. For example, [22] proposes an approach based on hill climbing techniques combined with analytic queuing models to guide the search for the best combination of configuration parameters of a multi-layered architecture hosting E-Commerce applications. Again, [23] studies a theoretical model that uses both load balancing and server scheduling when trying to maximize the profit of a hosting platform subject to multi-class SLAs.

Finally, while there is an extensive literature on request-based admission control, session-based admission control is much less well studied. However, nobody has studied the effects of combining admission control, resource allocation and economics when trying to model a commercial service provisioning system subject to QoS constraints. For example, [24] and [25] consider some economic models dealing with single jobs, but they focus on allocating server capacity only, while admission policies are not taken into account. Yet, revenues can be improved very significantly by imposing suitable conditions for accepting user demand. The most closely related work is perhaps the one described in [5]. In that study, charges, obligations, penalties and admission policies apply to single jobs, and thus that model cannot be applied to session-based traffic.

## VII. Conclusions and Future Work

This paper has discussed the design and implementation of a prototype service provisioning system, called SPIRE, that tries to improve the efficiency of service provisioning systems subject to QoS contracts. We have demonstrated that policy decisions such as server allocations and admission control can have a significant effect on the revenue. Moreover, those decisions are affected by the contractual obligations between clients and provider in relation to the QoS. SPIRE dynamically enforces SLAs by monitoring traffic parameters, income and expenditure, and by making dynamic decisions about server allocation and admission control. Three business models dealing with session-based traffic were introduced, and experiments have shown that they perform well under different traffic conditions.

Possible directions for future research include sharing a server among several types of services or expensive system reconfigurations, either in terms of money or time (Amazon EC2, for example, can take up to 10 minutes to launch a new instance). Also in order to further improve the efficiency of the available servers, a concurrency level higher than one could be used. Of course, since the SLAs are still in operation, it is not possible to change the concurrency level at random: instead, the same QoS level as if jobs were run alone should be maintained. Finally, the problem of designing dynamic policies that optimize performance and power consumption simultaneously is by no means solved yet.

## Acknowledgments

This work was carried out while the author was with Newcastle University. It was partly funded by British Telecom, under the research project QOSP (Quality of Service Provisioning). The author would like to thank Isi Mitrani for the useful discussions and insights, and Elisa Turrini, Manuel Mazzara, Dmytro Dyachuk and Ivona Brandic for the feedback on an earlier draft of this paper.